\documentclass[letterpaper,english,preprint, aps,nofootinbib]{revtex4-1}
\usepackage[T1]{fontenc}
\usepackage[latin9]{inputenc}
\usepackage{babel}
\usepackage{tipa}
\usepackage{tipx}
\usepackage{amsmath}
\usepackage{amssymb}
\usepackage{stmaryrd}
\usepackage[unicode=true,pdfusetitle,
 bookmarks=true,bookmarksnumbered=false,bookmarksopen=false,
 breaklinks=false,pdfborder={0 0 1},backref=false,colorlinks=false]
 {hyperref}

\makeatletter

\pdfpageheight\paperheight
\pdfpagewidth\paperwidth

\usepackage{etoolbox}
\usepackage{breqn}

\makeatletter
\let\cat@comma@active\@empty
\makeatother

\newcommand{\breqnoverloadothers}
{%
    \renewenvironment{equation}{\ignorespaces\begin{dmath}}{\end{dmath}\ignorespacesafterend}%
    \renewenvironment{equation*}{\ignorespaces\begin{dmath*}}{\end{dmath*}\ignorespacesafterend}%
    \renewenvironment{multline}{\ignorespaces\begin{dmath}}{\end{dmath}\ignorespacesafterend}%
    \renewenvironment{multline*}{\ignorespaces\begin{dmath*}}{\end{dmath*}\ignorespacesafterend}%

}

\newcommand\breqnundefineothers
{%
    \renewenvironment{equation}{}{}%
    \renewenvironment{equation*}{}{}%
    \renewenvironment{multline*}{}{}%

}

\AtBeginEnvironment{dmath}{\breqnundefineothers}
\AtBeginEnvironment{dmath*}{\breqnundefineothers}

\AtBeginEnvironment{dgroup}{\def\breqnundefineothers{}\breqnoverloadothers}
\AtBeginEnvironment{dgroup*}{\def\breqnundefineothers{}\breqnoverloadothers}

\makeatother

\begin{document}
\title{Semiclassical Dynamics of Hawking Radiation}
\author{David A. Lowe}
\email{Author to whom any correspondence should be addressed. lowe@brown.edu}

\affiliation{Physics Department, Brown University, Providence, RI 02912, USA.}
\author{L\textipa{\' a}rus Thorlacius}
\email{lth@hi.is}

\affiliation{Science Institute, University of Iceland, Dunhaga 3, 107 Reykjavík,
Iceland}
\begin{abstract}
We consider gravity in 3+1 spacetime dimensions coupled to $N$ scalar
matter fields in a semiclassical limit where $N\to\infty$. The dynamical
evolution of a black hole including the back-reaction of the Hawking
radiation on the metric is formulated as an initial-value problem.
The quantum stress-energy tensor is evaluated using a point-splitting
regularization along spacelike geodesics. To account for the quantum
entanglement of the matter fields, they are treated as a set of bilocal
collective fields defined on spacelike hypersurfaces. The resulting
semiclassical field equations include terms up to fourth order in
derivatives that can be treated in a perturbative $\hbar$ expansion.
The formulation we arrive at should be amenable to numerical simulation
of time-dependent semiclassical spacetime.
\end{abstract}
\maketitle

\section{Introduction\label{sec:Introduction}}

The semiclassical approximation in gravity allows us to consider quantum
effects, such as Hawking emission from black holes and the origin
of cosmological perturbations via inflation. Semiclassical considerations,
including Hawking's original calculation of black hole radiation~\citep{Hawking:1975vcx},
typically involve the quantization of matter fields in a classical
spacetime geometry that is obtained as a solution of Einstein's equations
without any quantum corrections. In principle, the back-reaction of
Hawking emission on the black hole metric can be incorporated by solving
a semiclassical Einstein equation of the form
\begin{equation}
G_{\mu\nu}=8\pi G_{N}\left\langle T_{\mu\nu}\right\rangle ,\label{eq:semieinstein}
\end{equation}
where the expectation value $\left\langle T_{\mu\nu}\right\rangle $
is of a suitably renormalized matter stress tensor in a quantum state,
$G_{\mu\nu}$ is the Einstein tensor and $G_{N}$ is Newton's constant.
In practice, there are technical and conceptual obstacles to overcome,
and it has remained a long-standing problem to even formulate a self-consistent
set of equations that incorporate the semiclassical back-reaction
in four spacetime dimensions. Reviews of semiclassical gravity can
be found in \citep{Birrell:1982ix,Wald:1984rg,Wald:1995yp} as well
as original papers such as \citep{Hawking:1975vcx,PhysRevD.14.870}.

In this paper, we revisit the back-reaction problem and present a
new approach for its solution. We couple Einstein gravity to matter
in the form of $N$ scalar fields and for the renormalization of the
matter stress-energy tensor we employ a variant of the geodesic point-splitting
method introduced by DeWitt \citep{DeWitt:1960fc,DeWitt:1964mxt}
and further developed by a number of authors including \citep{Christensen:1976vb,Adler:1976jx,Wald:1978pj,Candelas:1980zt,Brown:1986tj,Decanini:2005eg}.
Point-splitting provides an explicit coordinate invariant regularization
scheme that is well suited to an initial-value formulation, but its
detailed implementation is somewhat subtle, especially when the goal
is to have a flexible formalism that applies to generic dynamical
backgrounds. We adopt the minimal subtraction prescription of \citep{Brown:1986tj,Decanini:2005eg}
rather than subtracting a Hadamard elementary solution to define a
regularized matter stress tensor. The minimal subtraction only depends
on the local geometry, which is an important simplification when it
comes to implementation, but the resulting subtracted matter two-point
functions fail to satisfy the usual field equations. We show how to
circumvent this problem by working with bilocal collective fields,
formed from equal-time correlation functions of scalar fields inserted
at distinct spatial points. The bilocal fields are evolved forward
in time using standard Schwinger-Dyson equations and in the coincident
limit, they yield a renormalized stress-energy tensor for the right-hand
side of the semiclassical Einstein equation \eqref{eq:semieinstein}.
The resulting coupled dynamical equations govern the semi-classical
evolution of the metric and matter fields starting from general initial
data.\footnote{For consistency, the initial data must satisfy the semi-classical
constraint equations, but no further restrictions need to be imposed.} Our formalism will by construction reproduce various earlier results
obtained via point-splitting for static spacetime geometries with
special symmetries, but it can also be applied to fully dynamical
backgrounds that are not restricted to any particular symmetry. While
the main motivation for the present work comes from black hole theory,
the same methods can also be applied to study semiclassical back-reaction
in quantum cosmology.

The one-loop effective actions for generic quantum fields in curved
spacetime are non-local \citep{Barvinsky:1990up}, making it a challenge
to articulate a well-posed initial-value problem. In the simpler context
of two-dimensional dilaton gravity coupled to scalar fields, the corresponding
problem can be cast in terms of a non-local Polyakov action \citep{POLYAKOV1981207,Callan:1992rs}.
In that case, the effective action can be expressed in a local form
by choosing a conformal gauge for the two-dimensional metric. To our
knowledge, there is no choice of gauge that renders the four-dimensional
problem local, and it has been suggested the non-local character of
the effective action is essential to resolving the black hole information
problem \citep{Calmet:2021cip}. However, any semiclassical theory
that treats the spacetime metric as a classical field will lose information
to black holes and one must look beyond the semiclassical approximation
to restore unitarity, for instance via a holographic dual field theory.
In the present paper, our goal is restricted to formulating consistent
dynamical equations for semiclassical time evolution in four-dimensional
spacetime rather than considering the fate of quantum information.

An interesting approach explored in \citep{Levi:2015eea,Levi:2016quh,Zilberman:2019buh}
involves expressing the regularized stress tensor in frequency or
angular momentum space via sums over field modes. This framework again
is non-local in character, as the modes must be defined on the entire
spacetime. Nevertheless, the approach has been successful in computing
the complete renormalized stress-energy tensor in static and stationary
black hole backgrounds, extending the original work of Candelas \citep{Candelas:1980zt}
where some components of the point-split renormalized stress-energy
tensor were computed in a Schwarzschild background. 

The semiclassical theory is expected to simplify when one takes a
large $N$ limit, where $N$ is the number of species of quantum field.
In this paper, we will consider $N$ real-valued scalar fields $\phi_{\alpha}$
with mass $m$ and non-minimal coupling to gravity $\xi$ with an
action $S$ of the form
\begin{equation}
S=-\frac{1}{2}\int d^{4}x\sqrt{-g}\sum_{\alpha=1}^{N}\left(g^{\mu\nu}\partial_{\mu}\phi_{\alpha}\partial_{\nu}\phi_{\alpha}+m^{2}\phi_{\alpha}\phi_{\alpha}+\xi R\phi_{\alpha}\phi_{\alpha}\right).
\end{equation}
We have in mind scaling $\hbar\sim1/N$ so that fluctuations in the
matter fields are under control, while at the same time, they overwhelm
fluctuations in the metric, which we treat semiclassically in an $\hslash$
expansion. Important recent progress was made in \citep{Juarez-Aubry:2022qdp},
where semiclassical gravity was formulated as an initial-value problem,
subject to a number of conjectures. In the present work, we emphasize
the role of the large $N$ expansion in the number of matter fields,
which is crucial to make the semiclassical approximation well-defined.
In addition, we impose constraints on the space of states so that
a well-defined expansion emerges where the fluctuations in the metric
are $1/N$ suppressed. 

The leading-order semiclassical corrections to the Einstein equations
contain terms with up to four spacetime derivatives of the metric
\citep{Wald:1978pj}. This will in general lead to unphysical behavior
in solutions, but the problem can be sidestepped by treating the higher
derivative terms perturbatively in $1/N$, following \citep{Simon:1990jn,Parker:1993dk}.
The resulting semiclassical field equations will lead to a well-behaved
time evolution for smooth initial data, as long as the spacetime curvature
remains small compared to the Planck scale. Finding analytic solutions
in closed form is likely beyond reach, but the formalism provides
a jumping-off point for numerical computations of Hawking emission
from black holes in four spacetime dimensions with semiclassical back-reaction
included.

The remainder of the paper is organized as follows. In section II
we briefly review the traditional approach to general relativity as
an initial-value problem. In section III we outline the evaluation
of the stress-energy tensor in a $1/N$ expansion and in section IV
we introduce the scalar field degrees of freedom and their evolution
equations. These take the form of the usual classical local fields,
which satisfy the Klein-Gordon wave equation, as well as bilocal fields
which satisfy a set of Schwinger-Dyson equations. In section V we
show how the complete renormalized stress-energy tensor may be computed
in an arbitrary background using these scalar degrees of freedom together
with operator counter-terms needed to restore the diffeomorphism invariance
broken by the point-splitting regularization. In section VI we describe
how the evolution equations reduce in a variety of special limits
to results in the existing literature. We conclude in section VII.

\section{General Relativity as an Initial-Value Problem\label{sec:General-Relativity-as}}

In this section, we briefly review the formulation of the constrained
initial-value problem typically used in numerical general relativity
calculations, following \citep{York:1978gql} and a more recent review
\citep{Gourgoulhon:2007ue}. This results in a set of evolution equations
for the gravitational variables, with the classical matter stress
tensor as a source. Semiclassical evolution equations are then obtained
by including leading-order quantum corrections in the stress tensor.

We begin with a general metric written in terms of the lapse $N$
and shift vector $\beta^{i}$,\footnote{We also use the symbol $N$ to denote the number of species of scalar
fields. We presume the usage will be clear from the context.}
\begin{equation}
ds^{2}=g_{\mu\nu}dx^{\mu}dx^{\nu}=-N^{2}dt^{2}+\gamma_{ij}\left(dx^{i}+\beta^{i}dt\right)\left(dx^{j}+\beta^{j}dt\right),
\end{equation}
where $\mu,\nu=0,\cdots,3$ are spacetime indices and $i,j=1,2,3$
are purely spatial. A fixed $t$ slice is a spacelike hypersurface
$\Sigma$. 

The Riemann tensor can be decomposed into the intrinsic Riemann tensor
on $\Sigma$ and terms involving the extrinsic curvature $K_{ij}$
of $\Sigma$ embedded in spacetime. We take $n^{\mu}$ to be the timelike
unit normal to $\Sigma$, defined as
\begin{equation}
n_{\mu}=-N\frac{\partial t}{\partial x^{\mu}}\,.
\end{equation}
An orthogonal projector can be defined which projects into the tangent
space of $\Sigma$,
\begin{equation}
\gamma_{\beta}^{\alpha}=\delta_{\beta}^{\alpha}+n^{\alpha}n_{\beta}\,.
\end{equation}
The extrinsic curvature is defined with a minus sign convention, 
\begin{equation}
K_{\alpha\beta}=-\gamma_{\alpha}^{\mu}\gamma_{\beta}^{\nu}\nabla_{\mu}n_{\nu}\,.
\end{equation}
The induced metric has a unique Levi-Civita connection $D_{i}$ associated
with it. This in turn defines the intrinsic curvature tensor of the
hypersurface, which we denote by $R_{\,\,lij}^{k}$. The corresponding
Ricci scalar is also known as the Gaussian curvature of the surface.

The stress-energy tensor can likewise be decomposed into components
tangent to $\Sigma$ and components normal to $\Sigma$ using $n^{\mu}$
and the orthogonal projector 
\begin{equation}
E=T_{\mu\nu}n^{\mu}n^{\nu}\,,\quad p_{\alpha}=T_{\mu\nu}n^{\mu}\gamma_{\alpha}^{\nu}\,,\quad S_{\alpha\beta}=T_{\mu\nu}\gamma_{\alpha}^{\mu}\gamma_{\beta}^{\nu}\,.\label{eq:stress}
\end{equation}

Before writing the Einstein equations as an initial-value problem,
it is convenient to define a rescaled timelike normal,
\begin{equation}
m_{\mu}=Nn_{\mu},
\end{equation}
which is dual to $dt$. If we translate each point on $\Sigma$ by
$m^{\mu}\delta t$ then the value of the new time coordinate is simply
$t+\delta t$. 

The Einstein equations (in trace-reversed form) can then be written,
using Lie derivatives with respect to $m$, as
\begin{align}
\mathcal{L}_{m}\gamma_{ij} & =-2NK_{ij},\nonumber \\
\mathcal{L}_{m}K_{ij} & =-D_{i}D_{j}N+N\left(R_{ij}+KK_{ij}-2K_{ik}K_{j}^{k}+4\pi G_{N}\left(\left(S-E\right)\gamma_{ij}-2S_{ij}\right)\right),\label{eq:einstein}
\end{align}
together with the constraints
\begin{align}
R+K^{2}-K_{ij}K^{ij} & =16\pi G_{N}E\,,\nonumber \\
D_{j}K_{i}^{j}-D_{i}K & =8\pi G_{N}p_{i}\,.\label{eq:constraints}
\end{align}
If the stress-energy tensor satisfies $\nabla_{\mu}T^{\mu\nu}=0$,
then the Bianchi identities guarantee the constraints are satisfied
on subsequent timeslices. 

In the following, our goal will be to formulate an initial-value problem
to determine a semiclassical stress tensor $\left\langle T_{\mu\nu}\right\rangle $
which we can then insert into these equations. From the numerical
viewpoint, the BSSN approach \citep{nakamura,baumgarte} yields a
more numerically stable setup than the familiar ADM approach \citep{Arnowitt:1962hi}
described above. Both approaches are reviewed in \citep{Gourgoulhon:2007ue}
and it is straightforward to change to the BSSN variables once $\left\langle T_{\mu\nu}\right\rangle $
is known. 

\section{Semiclassical Expansion\label{sec:Semiclassical-Expansion}}

We will find that quantum fluctuations of $N$ scalar fields modify
the Einstein equations by terms up to fourth order in time derivatives.
Higher-derivative corrections also arise if we treat Einstein gravity
as an effective field theory, where we expect terms such as $R^{2}$
and $R_{\mu\nu}R^{\mu\nu}$ to appear in the effective action, with
coefficients to be matched to experiment. Terms of high derivative
order generically introduce spurious unphysical solutions to field
equations. To treat such terms, we will follow the general philosophy
advocated in \citep{Simon:1990jn,Parker:1993dk} where for physical
solutions of some more complete theory, there are constraints on the
solutions that appear when one expands the equations of motion in
a derivative expansion. 

More concretely, we will treat the higher derivative corrections as
contributions to the matter stress tensor and expand the resulting
$\left\langle T_{\mu\nu}\right\rangle $ as a power series in $\hbar$,
\begin{equation}
\left\langle T_{\mu\nu}\right\rangle =T_{\mu\nu}^{cl}+\sum_{n=1}^{\infty}\hbar^{n}\left\langle T_{\mu\nu}^{(n)}\right\rangle ,
\end{equation}
which can then be substituted directly into \eqref{eq:stress} to
obtain the semiclassical initial-value equations. We restrict to scalar
field states for which the equations of motion at zeroth order in
$\hbar$ reduce to the ordinary Einstein equations, which are second
order in time derivatives and subject to standard initial-value methods
as described in the previous section.\footnote{On this point, we differ with \citep{Juarez-Aubry:2022qdp}, who instead
allow for scalar field states corresponding to superpositions of distinct
classical states even at zeroth order. Such an approach will typically
not lead to a convergent $\hbar$ expansion.} The classical contribution $T_{\mu\nu}^{cl}$ is determined by the
scalar field one-point functions \begin{dmath}
\begin{equation}
T_{\mu\nu}^{cl}=\sum_{\alpha=1}^{N}\left(\left(1-2\xi\right)\partial_{\mu}\phi_{\alpha}^{cl}\partial_{\nu}\phi_{\alpha}^{cl}+\left(2\xi-\frac{1}{2}\right)g_{\mu\nu}g^{\lambda\rho}\partial_{\lambda}\phi_{\alpha}^{cl}\partial_{\rho}\phi_{\alpha}^{cl}+2\xi\phi_{\alpha}^{cl}\left(g_{\mu\nu}\square\phi_{\alpha}^{cl}-\nabla_{\mu}\nabla_{\nu}\phi_{\alpha}^{cl}\right)+\left(\xi G_{\mu\nu}-\frac{1}{2}g_{\mu\nu}m^{2}\right)\left(\phi_{\alpha}^{cl}\right)^{2}\right),
\end{equation}
\end{dmath}with $\phi_{\alpha}^{cl}\equiv\left\langle \phi_{\alpha}\right\rangle $.
The $O(\hslash)$ correction to $T_{\mu\nu}$ will be computed below. 

To proceed, we likewise make an expansion of the metric,
\begin{equation}
g_{\mu\nu}=g_{\mu\nu}^{cl}+\sum_{n=1}^{\infty}\hbar^{n}g_{\mu\nu}^{(n)}\,.
\end{equation}
At each time step, we then have in mind performing an iterative procedure
where we substitute in the $g_{\mu\nu}^{(n)}$ with $n<l$ into the
expression for $\left\langle T_{\mu\nu}^{(l)}\right\rangle $ and
using the Einstein equations with the corrected stress tensor, to
compute the next higher-order correction $g_{\mu\nu}^{(l)}$. Note
the higher time derivative terms of $g_{\mu\nu}^{(n)}$ may be computed
simply by taking time derivatives of the Einstein equations truncated
to order $n$. The Einstein equations are nonlinear so in practice,
one must take $\hbar$ sufficiently small that one achieves numerical
convergence with a truncated series. There is of course no guarantee
this expansion will converge for all times, but for black hole evolution,
we only expect non-convergence to become a problem once curvatures
become of order the Planck scale, near the evaporation endpoint.

\section{Semiclassical Equations of Motion\label{sec:Semiclassical-Equations}}

The semiclassical approximation then boils down to making the replacement
$T_{\mu\nu}\to\left\langle T_{\mu\nu}\right\rangle $ where the contribution
from the scalar fields is to be computed. The leading order contribution
to the stress-energy tensor from the classical scalars $\phi_{\alpha}^{cl}$
gives rise to classical solutions of general relativity, including
black holes formed in gravitational collapse, but in order to generate
the Hawking effect we need to include higher order contributions that
take into account the quantum entanglement of the scalar fields. 

With four-dimensional spacetime, our strategy will be to define spatially
bilocal collective fields on the spacelike surface $\Sigma_{t}$ in
terms of equal-time correlation functions of scalar fields inserted
at distinct spatial points $\vec{x}\neq\vec{x}'$ ,
\begin{align}
\Psi(\vec{x},\vec{x}';t) & =\frac{1}{N}\sum_{\alpha}\left\langle \phi_{\alpha}(\vec{x},t)\phi_{\alpha}(\vec{x}',t)\right\rangle ,\nonumber \\
\Upsilon(\vec{x},\vec{x}';t) & =\frac{1}{N}\sum_{\alpha}\left\langle \dot{\phi}_{\alpha}(\vec{x},t)\phi_{\alpha}(\vec{x}',t)\right\rangle ,\label{eq:collective}\\
\Omega(\vec{x},\vec{x}';t) & =\frac{1}{N}\sum_{\alpha}\left\langle \dot{\phi}_{\alpha}(\vec{x},t)\dot{\phi}_{\alpha}(\vec{x}',t)\right\rangle ,\nonumber 
\end{align}
where 
\begin{equation}
\dot{\phi}_{\alpha}=\mathcal{L}_{m}\phi_{\alpha},
\end{equation}
and treat these collective fields as dynamical variables. As with
the metric and $T_{\mu\nu}$ each of these collective fields will
have an $\hbar$ expansion as per the previous section. We have in
mind taking a large $N$ limit, and scaling $\hbar\sim1/N$ so that
fluctuations in the matter fields are suppressed, while at the same
time overwhelm the fluctuations in the metric, which we treat semiclassically.

An alternative approach would be to integrate out the matter fields
and define an effective action. However, we would like our formalism
to extend to massless matter fields, possibly conformally coupled,
and in this case, the effective action would be extremely non-local.
One would need to come up with a clever gauge-fixing scheme to allow
a standard initial-value problem to be formulated, analogous to conformal
gauge in the two-dimensional case. 

The expectation values in \eqref{eq:collective} involve unrenormalized
products of the fields. We can evolve these variables forward in time
using the Schwinger-Dyson equations. Up to contact terms, these may
be derived by inserting the scalar field equation of motion,
\begin{equation}
\left(\square_{x}-m^{2}-\xi R\right)\phi(x)=0\,,\label{eq:kleing}
\end{equation}
into the correlators. We will employ point splitting to regularize
divergences and contact terms will not play a role. To see how the
Schwinger-Dyson equations generate a closed system of equations amongst
the collective fields, we apply a time derivative to the expressions
in \eqref{eq:collective},
\begin{align}
\frac{\partial}{\partial t}\Psi(\vec{x},\vec{x}';t) & =\Upsilon(\vec{x},\vec{x}';t)+\Upsilon(\vec{x}',\vec{x};t),\nonumber \\
\frac{\partial}{\partial t}\Upsilon(\vec{x},\vec{x}';t) & =\Omega(\vec{x},\vec{x}';t)+\frac{1}{N}\sum_{\alpha}\left\langle \ddot{\phi}_{\alpha}(\vec{x},t)\phi_{\alpha}(\vec{x}',t)\right\rangle ,\label{eq:collective-1}\\
\frac{\partial}{\partial t}\Omega(\vec{x},\vec{x}';t) & =\frac{1}{N}\sum\left(\left\langle \ddot{\phi}_{\alpha}(\vec{x},t)\dot{\phi}_{\alpha}(\vec{x}',t)\right\rangle +\left\langle \dot{\phi}_{\alpha}(\vec{x},t)\ddot{\phi}_{\alpha}(\vec{x}',t)\right\rangle \right),\nonumber 
\end{align}
and then use the field equation \eqref{eq:kleing} to eliminate the
$\ddot{\phi}_{\alpha}(\vec{x},t)$ and $\ddot{\phi}_{\alpha}(\vec{x}',t)$
inside the correlation functions in favor of terms with at most one
time derivative acting on $\phi_{\alpha}(\vec{x},t)$ and $\phi_{\alpha}(\vec{x'},t)$,
respectively. As a result, the right-hand sides of all three equations
in \eqref{eq:collective-1} can be expressed as a sum of terms involving
the collective fields themselves and their spatial derivatives with
coefficients that depend on the spacetime metric. Thus the collective
fields satisfy a set of linear partial differential equations, of
first order in time derivatives, that generate the time evolution
of the matter fields. The scalar fields are coupled to gravity and
the full set of coupled evolution equations also includes the semiclassical
Einstein equations that we discuss in the next section.

It is important to use the unrenormalized correlators since, as we
will see, the subtracted correlators will only satisfy \eqref{eq:kleing}
up to source terms that depend on the non-local subtraction term.
In general, we will only need to compute this subtraction term in
the coincident limit, in order to find the stress-energy tensor. By
working with the bilocal collective fields we avoid having to consider
subtraction at finite spatial distance.

The one-point functions of the scalar fields $\phi_{\alpha}^{cl}=\left\langle \phi_{\alpha}\right\rangle $
must be treated as an independent set of local degrees of freedom
in order to compute the classical contribution $T_{\mu\nu}^{cl}$.
The one-point functions are determined by solving the standard Klein-Gordon
equation in a curved spacetime background,

\begin{equation}
\left(\square_{x}-m^{2}-\xi R\right)\phi_{\alpha}^{cl}(x)=0\,.
\end{equation}

\section{Stress-Energy Tensor via Point-Splitting Regularization}

It will simplify the discussion to temporarily switch to a covariant
notation as we discuss the regularization of the stress tensor using
point-splitting. We will then be free to drop back to the 3+1 notation
of section \eqref{sec:General-Relativity-as}. For the moment we will
drop the classical contributions and add them back in section \ref{subsec:Renormalized-Stress-Energy}.
To compute $\left\langle T_{\mu\nu}\right\rangle $ we use point-splitting
regularization to define \citep{DeWitt:1960fc,Decanini:2005eg}
\begin{align}
\left\langle T_{\mu\nu}\right\rangle  & =\lim_{x\to x',t\to t'}\left((1-2\xi)\bar{g}_{\nu}^{\,\nu'}\nabla_{\mu}\nabla{}_{\nu'}+\left(2\xi-\frac{1}{2}\right)g_{\mu\nu}\bar{g}^{\rho\lambda'}\nabla_{\rho}\nabla_{\lambda'}-\frac{1}{2}g_{\mu\nu}m^{2}\right.\nonumber \\
 & \left.+2\xi\left(g_{\mu\nu}\nabla_{\rho}\nabla^{\rho}-\bar{g}_{\mu}^{\,\mu'}\bar{g}_{\nu}^{\,\nu'}\nabla_{\mu'}\nabla_{\nu'}\right)+\xi G_{\mu\nu}\right)\sum_{\alpha=1}^{N}\left\langle \phi_{\alpha}(x)\phi_{\alpha}(x')\right\rangle ,\label{eq:stressenergy}
\end{align}
where, for our purposes, the events $x$ and $x'$ are connected by
a spacelike geodesic. Before taking limits, the object on the right-hand
side is a tensor in $x$ and a scalar in $x'$. In the language of
\citep{DeWitt:1960fc} this is a bitensor, and it is necessary to
use the bivector of parallel transport $\bar{g}_{\mu}^{\nu'}$ to
transport vectors from $x'$ to $x$ in order to develop a covariant
expansion of the point-split stress-energy tensor around $x$. 

As discussed by Wald \citep{Wald:1978pj}, following work of Adler
\emph{et al.} \citep{Adler:1976jx}, the point-splitting regularization
necessitates the addition of local curvature counter-terms to $\left\langle T_{\mu\nu}\right\rangle $
to restore the relation $\nabla^{\mu}\left\langle T_{\mu\nu}\right\rangle =0$.
The added terms cannot be obtained from any local effective action,
and they generate a quantum anomaly in the trace of the stress-energy
tensor. Once we have constructed a subtracted stress-energy tensor,
including any counter-terms that arise in the regularization procedure,
we may project onto the hypersurface $\Sigma$ using \eqref{eq:stress}
and insert the resulting semiclassical expressions into the Einstein
equations \ref{eq:einstein} and \ref{eq:constraints}. 

Let us briefly review the approach of \citep{Wald:1978pj,Adler:1976jx}
to define a renormalized stress-energy tensor. The idea is to replace
$\left\langle \phi_{\alpha}(x)\phi_{\alpha}(x')\right\rangle $ in
\eqref{eq:stressenergy} by
\begin{equation}
\left\langle \phi_{\alpha}(x)\phi_{\alpha}(x')\right\rangle _{B}=\left\langle \phi_{\alpha}(x)\phi_{\alpha}(x')\right\rangle -\left\langle \phi_{\alpha}(x)\phi_{\alpha}(x')\right\rangle _{L}
\end{equation}
where $\left\langle \phi(x,t)\phi(x',t')\right\rangle _{L}$ is a
Hadamard elementary solution. This is a natural way to employ the
point-splitting regularization as introduced in \citep{DeWitt:1960fc}
and further studied by \citep{Adler:1976jx}. Here $B$ denotes the
part of the two-point function that is state-dependent, while $L$
denotes the local part of the two-point function that is entirely
determined by the exact metric. Before we proceed to define these
$L$ and $B$ terms, it is helpful to review some facts about the
Hadamard expansion.

\subsection{Hadamard Expansion}

Consider a correlator $G(x,x')$ that solves the scalar wave equation
in the $(x,t)$ variables, with the light-cone singularities of the
object made explicit,
\begin{equation}
G(x,x')=\frac{1}{2\left(2\pi\right)^{2}}\left(\frac{U(x,x')}{\sigma}+V(x,x')\log\sigma+W(x,x')\right).\label{eq:hadamard}
\end{equation}
Here $\sigma(x,x')$ is half the square of the geodesic distance between
$x$ and $x'$. It is a biscalar quantity and satisfies the following
relation,
\begin{equation}
g^{\mu\nu}\sigma_{;\mu}\sigma_{;\nu}=2\sigma\,.\label{eq:sigmarelation}
\end{equation}
The biscalar functions $U,V,W$ are regular as $x'\rightarrow x$
and it is useful to express $V,W$ as power series expansions,
\begin{align}
V(x,x') & =\sum_{n=0}^{\infty}V_{n}(x,x')\sigma^{n}\,,\nonumber \\
W(x,x') & =\sum_{n=0}^{\infty}W_{n}(x,x')\sigma^{n}\,.\label{eq:vwexpansions}
\end{align}
To find $U,V$and $W$ we demand that \eqref{eq:hadamard} satisfies
the scalar wave equation in $x$,
\begin{equation}
\left(\square_{x}-m^{2}-\xi R\right)G(x,x')=0\,,\label{eq:hadeqn}
\end{equation}
and consider the left-hand side of the equation term by term in a
small geodetic distance expansion, using \eqref{eq:sigmarelation}
to simplify expressions. Setting the overall coefficient of the leading
$\sigma^{-2}$ term to zero yields
\begin{equation}
(2U_{;\mu}-U\Delta^{-1}\Delta_{;\mu})\sigma^{;\mu}=0\,.\label{eq:ueqn}
\end{equation}
The biscalar $\Delta$ is the Van Vleck-Morette determinant \citep{DeWitt:1960fc},
\begin{equation}
\Delta(x,x')=-(-g(x))^{-1/2}(-g(x'))^{-1/2}\det\left(-\sigma(x,x')_{;\mu\nu}\right),
\end{equation}
and is determined completely by the metric. Equation \eqref{eq:ueqn}
may be solved by integrating along any geodesic emanating from point
$x$. With the boundary condition that $U(x,x)=1,$ the unique solution
is $U=\Delta^{1/2}$.

Setting the $\log\sigma$ term to zero in \eqref{eq:hadeqn} implies
the function $V$ satisfies the wave equation,
\begin{equation}
\left(\square_{x}-m^{2}-\xi R\right)V(x,x')=0\,,\label{eq:veqn}
\end{equation}
subject to the following condition on the leading term in the short
distance expansion \eqref{eq:vwexpansions},
\begin{equation}
2V_{0}+2V_{0;\mu}\sigma^{;\mu}-2V_{0}\Delta^{-1/2}\Delta_{;\mu}^{1/2}\sigma^{;\mu}+\left(\square_{x}-m^{2}-\xi R\right)\Delta^{1/2}=0\,,
\end{equation}
which comes from the $1/\sigma$ coefficient in \eqref{eq:hadeqn}.
This implies $V_{0}$ can be integrated along geodesics and be determined
in terms of geometric quantities, which in turn determines the other
coefficients $V_{n}$ and the whole of $V(x,x')$ in terms of geometric
quantities. It can be shown this function is symmetric in $x$ and
$x'$.

Finally, the function $W(x,x')$ satisfies an inhomogeneous wave equation
of the form,
\begin{equation}
\sigma\left(\square_{x}-m^{2}-\xi R\right)W=-\left(\square_{x}-m^{2}-\xi R\right)\Delta^{1/2}-2V-2V_{;\mu}\sigma^{;\mu}+2V\Delta^{-1/2}\Delta_{;\mu}^{1/2}\sigma^{;\mu}.\label{eq:weqn}
\end{equation}
The full solution $W(x,x')$ is only determined up to an arbitrary
regular solution $W_{0}(x,x')$ of the corresponding homogenous wave
equation, which needs to be specified. This is where the state dependence
of the solution makes its appearance. To specify a Hadamard elementary
solution, $G_{L}(x,x')$ we set $W_{0}=0$, and then the resulting
function $W$ is determined purely geometrically. Henceforth, we will
denote the $W$ solution with $W_{0}=0$ as $W_{L}$. 

The state-dependent contribution $G_{B}(x,x')$ will satisfy the same
equations but with $W_{0}\neq0$ in general. In that case $U$ and
$V$ will be the same as before, but we denote $W=W_{B}$ for this
solution.

As it turns out \citep{Wald:1978pj}, it is not true that $W_{L}$
(or $W_{B}$) is symmetric in $x$ and $x'$ and this makes it necessary
to add counterterms to the stress-energy tensor. In fact, the wave
equation that $G_{L}$ satisfies in the $x'$ variable contains an
inhomogeneous term, 
\begin{align}
\left(\square_{x'}-m^{2}-\xi R\right)G_{L}(x,x') & =\frac{1}{2\left(2\pi\right)^{2}}\left(\left(\left(\square_{x'}-m^{2}-\xi R\right)W_{L}(x,x')\right)\right.\nonumber \\
 & \left.-\left(\square_{x}-m^{2}-\xi R\right)W_{L}(x',x)\right),
\end{align}
and is a slight generalization of equation (20) of \citep{Wald:1978pj}.
In the limit $x\to x'$, only the term $W=\sigma W_{1}$ survives
when one computes $\left\langle T_{\mu\nu}\right\rangle $, and this
may be solved for in terms of local geometric quantities. The upshot
is a non-vanishing contribution to $\nabla^{\mu}\left\langle T_{\mu\nu}\right\rangle $
which must be canceled by the addition of counterterms to $\left\langle T_{\mu\nu}\right\rangle $.
These counterterms are needed to restore the diffeomorphism invariance
broken by the point-splitting regulator. 

\subsection{Renormalized Stress-Energy Tensor\label{subsec:Renormalized-Stress-Energy}}

A slight modification of Wald's procedure was introduced in \citep{Brown:1986tj}
and generalized to the case of a massive non-minimally coupled scalar
in \citep{Decanini:2005eg}. Here one simply subtracts only the $1/\sigma$
and $\log\sigma$ terms in \eqref{eq:hadamard} and leaves out the
$W$ term. The modified procedure has the advantage that it maintains
symmetry under $x\leftrightarrow x'$ and that we avoid having to
compute the $W$ term. However, the price one pays for this is that
the subtraction term no longer satisfies the scalar wave equation
in $x$ or in $x'$. The procedure is sufficient to compute the finite
contributions to the expectation value of the stress-energy tensor,
and as discussed in Section \ref{sec:Semiclassical-Equations} above,
the failure of the subtraction term to satisfy the wave equation can
be sidestepped by working with the bilocal collective fields as the
dynamical variables of the matter sector.

The stress-energy tensor then takes the form\begin{dmath}
\begin{equation}
\left\langle T_{\mu\nu}\right\rangle =T_{\mu\nu}^{classical}+\left\langle T_{\mu\nu}^{bilocal}\right\rangle +\frac{N\hbar}{4\pi^{2}}g_{\mu\nu}\left(\frac{1}{8}m^{4}+\frac{1}{4}\left(\xi-\frac{1}{6}\right)m^{2}R-\frac{1}{24}\left(\xi-\frac{1}{5}\right)\square R+\frac{1}{8}\left(\xi-\frac{1}{6}\right)^{2}R^{2}-\frac{1}{720}R_{\sigma\tau}R^{\sigma\tau}+\frac{1}{720}R_{\sigma\tau\lambda\rho}R^{\sigma\tau\lambda\rho}\right)+\alpha\left(2\nabla_{\mu}\nabla_{\nu}R-2R\,R_{\mu\nu}+g_{\mu\nu}\left(\frac{1}{2}R^{2}-2\square R\right)\right)+\beta\left(R_{\mu\nu}-\square R_{\mu\nu}-2R^{\lambda\rho}R_{\mu\lambda\nu\rho}+\frac{1}{2}g_{\mu\nu}\left(R_{\sigma\tau}R^{\sigma\tau}-\square R\right)\right),\label{eq:fullstress}
\end{equation}
\end{dmath}where we have allowed for contributions that would arise
from local terms in the gravitational effective action of the form
$\alpha\int d^{4}x\sqrt{-g}R^{2}$ and $\beta\int d^{4}x\sqrt{-g}R_{\mu\nu}R^{\mu\nu}$
where $\alpha,\beta$ are order $N\hbar$ constants. They may, of
course, be simply set to zero, if one wishes to consider pure Einstein
gravity coupled to scalars, but in practice, such terms can arise
in an effective action and their coefficients should ultimately be
determined by experiment. We do not consider addition of $\int d^{4}x\sqrt{-g}R_{\mu\nu\lambda\rho}R^{\mu\nu\lambda\rho}$
since it may be combined with a particular linear combination of the
other two terms to yield the Gauss-Bonnet term, which is a total derivative. 

The contribution of the bilocal fields $\left\langle T_{\mu\nu}^{bilocal}\right\rangle $
enters at order $N\hbar$ and arises from evaluating \eqref{eq:stressenergy}
with the two-point function replaced by the subtracted two-point function,
\begin{equation}
\left\langle \phi_{\alpha}(x)\phi_{\alpha}(x')\right\rangle \rightarrow\left\langle \phi_{\alpha}(x)\phi_{\alpha}(x')\right\rangle -\frac{1}{2\left(2\pi\right)^{2}}\left(\frac{U(x,x')}{\sigma}+V(x,x')\log\sigma\right).\label{eq:subtraction}
\end{equation}
This gives rise to an expression that may then be expanded in terms
of our collective fields \eqref{eq:collective} defined on a spacelike
surface $\Sigma_{t}$, provided one uses the scalar field equation
\eqref{eq:kleing} inside the two-point correlators to eliminate terms
where two time derivatives act on a single scalar field, before taking
the equal-time limit $t'\rightarrow t$. The subtracted two-point
function in \eqref{eq:subtraction} is finite in the coincident limit
by construction and this guarantees that the final result for $\left\langle T_{\mu\nu}^{bilocal}\right\rangle $
is regular and can be included on the right-hand side of the semiclassical
Einstein equations \eqref{eq:semieinstein}.

The terms that generate the conformal anomaly are inside the parenthesis
multiplied by an explicit factor of $N\hslash$ in \ref{eq:fullstress}
and it is these terms combined with $\left\langle T_{\mu\nu}^{bilocal}\right\rangle $
that will give rise to Hawking radiation in a black hole background. 

A renormalization energy scale $M$ enters in the $\log\sigma$ term
in \eqref{eq:subtraction}. This may be redefined by $\log\sigma\to\log M^{2}\sigma$,
corresponding to choosing renormalization conditions for the scalar
fields. In turn, this leads to a dependence of the stress-energy tensor
on the choice of $M$. The formula \eqref{eq:fullstress} may then
be shifted by the terms \citep{Decanini:2005eg}\begin{dmath}
\begin{equation}
\left\langle T_{\mu\nu}\right\rangle \to\left\langle T_{\mu\nu}\right\rangle +\frac{\log M^{2}}{2\left(2\pi\right)^{2}}\left(g_{\mu\nu}\left(\frac{1}{8}m^{4}+\frac{1}{4}\left(\xi-\frac{1}{6}\right)m^{2}R-\frac{1}{2}\left(\xi^{2}-\frac{1}{3}\xi+\frac{1}{40}\right)\boxempty R+\frac{1}{8}\left(\xi-\frac{1}{6}\right)^{2}R^{2}-\frac{1}{720}R_{\lambda\rho}R^{\lambda\rho}+\frac{1}{720}R_{\sigma\tau\lambda\rho}R^{\sigma\tau\lambda\rho}\right)-\frac{1}{2}\left(\xi-\frac{1}{6}\right)m^{2}R_{\mu\nu}+\frac{1}{2}\left(\xi^{2}-\frac{1}{3}\xi+\frac{1}{30}\right)R_{;\mu\nu}-\frac{1}{120}\boxempty R_{\mu\nu}-\frac{1}{2}\left(\xi-\frac{1}{6}\right)^{2}RR_{\mu\nu}+\frac{1}{90}R_{\ \mu}^{\lambda}R_{\lambda\nu}-\frac{1}{180}R^{\lambda\rho}R_{\lambda\mu\rho\nu}-\frac{1}{180}R_{\ \ \ \mu}^{\lambda\sigma\tau}R_{\lambda\sigma\tau\nu}\right)\,.\label{eq:shift}
\end{equation}
\end{dmath}We note this shift ambiguity has vanishing trace in the
conformally coupled case, $m^{2}=0$ and $\xi=1/6$, so does not change
the expression for the trace anomaly.

The semiclassical stress-energy tensor in \eqref{eq:fullstress} includes
terms involving up to four time derivatives. Our strategy for dealing
with such terms will follow earlier work by \citep{Simon:1990jn,Parker:1993dk},
and we will perform an expansion for small $\hbar N$. From the numerical
viewpoint, we may treat the equations of motion as in section \ref{sec:General-Relativity-as}
in the standard way, however, at each time step, we must, in addition,
perform iterations to calculate the higher order in $\hbar N$ terms
in the metric. A similar approach has been advocated by \citep{Juarez-Aubry:2022qdp}.
We impose stronger constraints on the scalar field state in line with
the $1/N$ expansion, to ensure the $\hbar N$ expansion is uniform. 

\subsection{Initial-Value Problem for Coupled System}

Our goal is now to couple the fourth-order stress tensor \eqref{eq:fullstress}
to the usual second-order Einstein equations as set up in section
\ref{sec:General-Relativity-as}. The fourth-order local geometric
quantities may be handled according to the perturbative method described
in section \ref{sec:Semiclassical-Expansion}. The bilocal term may
be evaluated by substituting the set of collective fields \eqref{eq:collective}
with the subtraction term into the expression for the stress-energy
tensor \eqref{eq:stressenergy}.

With these expressions completed, it is worth summarizing our approach
to evolving the semiclassical dynamics by considering the formation
and subsequent evaporation of a black hole. In this case, the initial
data for the scalars and the metric is most easily set up by sending
a null shock from infinity into empty spacetime. The classical metric
is obtained by patching together a flat interior solution and a Schwarzschild
exterior, with suitable matching conditions across the null shockwave.
With the shock crossing the initial timeslice in the asymptotic region,
good approximations for the local and the bilocal fields are known
and can be used to set up the initial time step. The Schwinger-Dyson
equations may be used to generate subsequent time steps in the bilocal
collective fields and the Klein-Gordon equation for the classical
local fields. Likewise, at order $\hbar^{0}$ one's favorite 3+1 evolution
scheme may be used to obtain time derivatives of $\gamma_{ij}$ and
$\dot{\gamma}_{ij}$ with only the classical contribution to $\left\langle T_{\mu\nu}\right\rangle $
included. One may then compute higher order in $\hbar$ corrections
to these values by inserting the lower order solution into the $\hbar$
corrections to $\left\langle T_{\mu\nu}\right\rangle $ and iterating
until the time derivatives of $\gamma_{ij}$ and $\dot{\gamma}_{ij}$
converge. The point-splitting procedure for defining a renormalized
stress-energy tensor of the matter fields involves expanding various
quantities at short geodesic separation and implementing a subtraction
scheme. These expansions must refer to the coordinate system being
used in the actual numerical evolution, and they have to be carried
out to sufficiently high order to ensure the subtraction gives the
correct finite results. For completeness, we include the relevant
expansions for a general coordinate system in the Appendix.

At this point, one has a numerical approximation to the time derivatives
of the exact $\gamma_{ij}$ and $\dot{\gamma}_{ij}$ and the timestep
is complete. It is reasonable to expect the iteration generating the
$\hbar$ expansion to converge away from large spacetime curvatures,
assuming a nondegenerate choice of coordinates. In particular, we
expect the method to be applicable for smooth initial data describing
matter undergoing gravitational collapse to a macroscopic black hole.
For general non-vacuum initial data, it is important to ensure the
initial data satisfies the constraint equations and one must solve
an elliptic partial differential equation on the spacelike hypersurface
to enforce this. As described above, this solution should be treated
perturbatively in an $\hbar$ expansion, with the higher derivative
terms only appearing as small corrections to the leading order solution.
Such a procedure should converge away from regions with spacetime
singularities. An initial state with a spacetime singularity on the
initial slice is problematic in our approach not only because the
$\hbar$ expansion breaks down near a singularity, but because the
non-local entanglement captured by the bilocal fields cannot be reliably
determined on such a time slice. 

It is therefore important to choose an initial time slice that avoids
spacetime regions of strong curvature and implement a numerical scheme
for the detection and excision of curvature singularities in the subsequent
dynamical evolution. For general initial data there are many potential
issues that we will not attempt to resolve here, but for spherically
symmetric gravitational collapse we expect qualitatively similar results
to previous studies of two-dimensional dilaton gravity coupled to
matter \citep{Callan:1992rs,Russo:1992ax}, which have also been generalized
to include the spherically symmetric reduction of Einstein gravity
(though coupled to exotic matter from the 4d perspective) \citep{Lowe:1992ed}.
In the two-dimensional models a spacelike curvature singularity develops
in the black hole interior at which the semiclassical equations break
down. The singularity eventually intersects the receding apparent
horizon. The semiclassical theory cannot be extended to the causal
future of this endpoint but, importantly, almost all the Hawking radiation
from the black hole is emitted before the endpoint is reached.

Returning to four spacetime dimensions, one could choose for the initial
time slice a so-called nice slice, in the language of \citep{Lowe:1995ac},
which crosses the initial apparent horizon but avoids the strong curvature
region in the black hole interior. A possible singularity excision
scheme is to evaluate a local curvature invariant, such as the Kretschmann
scalar $K=R_{\mu\nu\lambda\sigma}R^{\mu\nu\lambda\sigma}$, on each
time slice and remove the causal future of any point where this quantity
exceeds a preset (large) value. This procedure will excise the high-curvature
region near the singularity from the geometry and, as long as the
cutoff value imposed on the curvature is sufficiently far below the
Planck scale to ensure convergence of the $\hbar$ expansion discussed
above, we expect the semiclassical solution to deform in a continuous
way to a static black hole (plus damped quasinormal modes) as $\hbar\to0$
on the spacetime region where the numerical solution has been obtained.
In other words, there should be a sense in which the isolated black
hole can be approximated at each timestep as a sequence of stationary
solutions (as suggested in Hawking's original paper \citep{Hawking:1975vcx})
together with a cloud of outgoing radiation. The semiclassical evolution
will terminate when the Hawking temperature of the remaining black
hole approaches the cutoff scale but at this point almost all the
Hawking radiation will already have been emitted.

The semiclassical approximation should be valid for the computation
of local quantities such as the expectation value of the stress energy
tensor, which in turn uniquely determines the time evolution of the
metric. On the other hand, non-local quantities such as von Neumann
entropy, or local quantities with the number of operator insertions
of order the black hole entropy, are not necessarily well-approximated
in this approach (see for example \citep{Lowe:2022cne} for a discussion
of these issues). Such observables are needed to address issues related
to the black hole information paradox. In the present approach we
do not expect to be able to compute these quantities reliably. However
within the semiclassical approximation we recover the usual information
loss scenario. We expect the answer to be qualitatively similar to
soluble models like \citep{Russo:1992ax} where asymptotic charges
such as energy are conserved, but all other quantum information is
lost.\footnote{This conclusion is not expected to change if one replaces the $N$
scalar fields with fields with nontrivial spin.} Nevertheless, the solution does reliably determine the semiclassical
spacetime, and one must go beyond semiclassical methods to determine
the fate of quantum information.

\section{Connections to Previous Work}

The main result of this paper is a self-consistent set of equations
governing the semiclassical evolution of general four-dimensional
spacetimes in a large $N$ limit, where $N$ is the number of scalar
fields. The quantum properties of the scalar fields are dealt with
in this approximation by treating the scalar field two-point function
as a dynamical bilocal variable. The state of the scalar field determines
the initial data for this bilocal variable, and evolution is determined
by solving the Schwinger-Dyson equations. Our evolution equations
allow for completely arbitrary choices of initial data for the metric
and bilocal variables, subject to the constraint equations, but in
special limits the equations simplify and one can make contact with
various previous work. 

\subsection{Hartle-Hawking Vacuum}

This is perhaps the easiest limit to consider, where naively a static
metric is supported by ingoing and outgoing thermal fluxes but this
is undermined by instability when semiclassical backreaction on the
spacetime metric is taken into account. For simplicity, we focus on
the non-rotating case, though analogous results are known for Kerr
black holes. At order $\hbar^{0}$ the metric is simply the static
Schwarzschild solution and the scalar field state in this vacuum leads
to a thermal two-point function with temperature equal to the Hawking
temperature. This sets the initial data for the metric (at order $\hbar^{0}$)
and for the bilocal variables. One may then proceed to compute the
renormalized stress-energy tensor \eqref{eq:fullstress}. At leading
order this can be achieved via sums over field modes defined on the
classical Schwarzschild background as in \citep{Candelas:1980zt,candelas,Levi:2015eea,Levi:2016quh}.
The stress-energy tensor obtained by these authors for the Hartle-Hawking
vacuum is regular at the event horizon and has a uniform density of
thermal radiation asymptotically far from the black hole.

The next step is to solve for the order $\hbar$ correction to the
metric, using the leading order stress tensor as source but this will
in general run into Jeans instabilities due to the finite thermal
energy density in the asymptotic region.\footnote{Finite energy density can be avoided by introducing charge or rotation
and considering the extremal limit where the Hawking temperature vanishes,
but this is not a generic black hole.} In practice, this leads to the Unruh vacuum as being of more physical
interest, due to its finite ADM mass.

In the special limit where one keeps the $\alpha,\beta$ terms in
\eqref{eq:fullstress} but discards the scalar-induced terms (proportional
to $N$) there is no asymptotic energy density and in this case one
will reproduce the quantum corrections to the black hole entropy and
temperature found by Wald \citep{Wald:1993nt}.

\subsection{Unruh Vacuum}

One may proceed in an analogous way for the Unruh vacuum, again taking
Schwarzschild as the initial metric at order $\hbar^{0}$ but choosing
a different vacuum state for the ingoing and outgoing scalar modes,
as described in detail in \citep{PhysRevD.14.870}. The scalar field
two-point function can be written as an expansion in terms of the
scalar mode functions in the Schwarzschild background, as made explicit
in \citep{Candelas:1980zt}. Again, one may proceed to compute the
induced stress-energy tensor using \eqref{eq:fullstress} reproducing
results of \citep{Candelas:1980zt,candelas,Levi:2015eea,Levi:2016quh}.
In this case, solving the metric correction at order $\hbar^{1}$
leads to a time-dependent metric, that requires numerical methods
to solve for, which are beyond the scope of the present paper, but
we hope to return to in future work.

\subsection{Boulware Vacuum}

One may take ingoing and outgoing scalar modes in the same vacuum
state but identify this as the state that annihilates positive frequency
modes with respect to time at infinity. The metric is taken to be
the static Schwarzschild metric at order $\hbar^{0}$. Upon computing
the induced stress-energy tensor on the horizon using \eqref{eq:fullstress},
one finds a divergence \citep{Candelas:1980zt} in a freely falling
reference frame that will lead to a curvature singularity in the order
$\hbar$ correction to the metric. The resulting semiclassical solution
is thus singular, in the sense that it violates weak cosmic censorship.

\subsection{Flat Spacetime}

As a consistency check, we verify that flat spacetime is an exact
solution of our full semiclassical equations of motion. The bilocal
fields are then identified with the known flat space scalar two-point
correlator (and time-derivatives thereof). For finite $m^{2}$ a nontrivial
cosmological constant is typically induced, as seen by the order $m^{4}$
term in \eqref{eq:fullstress}. One may, however, choose a renormalization
condition to cancel this at order $\hbar$ using the shift freedom
in \eqref{eq:shift} to eliminate the $m^{4}$ term. All remaining
terms in \eqref{eq:semieinstein} involve the spacetime curvature
and are manifestly zero in flat spacetime.

\section{Conclusions}

We have formulated a set of coupled field equations for scalar matter
and gravity in four spacetime dimensions that include semiclassical
back-reaction effects. These equations can be used to study the time
evolution of a black hole emitting Hawking radiation as an initial-value
problem. We apply point splitting along spacelike geodesics and the
counterterm needed to compute the renormalized stress-energy tensor
is developed as an expansion in a general coordinate system. The scalar
field dynamics involves a set of bilocal fields, which are functions
of a single time, but two space points on a given spacelike hypersurface.
The bilocal fields are necessary to capture the quantum entanglement
of the scalar fields, in a large $N$ approximation. The price one
pays for the relatively simple dynamical evolution of the bilocal
fields is that they become singular (in a prescribed way) in the coincident
limit, but the divergences can be handled by a subtraction procedure.
With these ingredients in place, one may then insert the renormalized
stress-energy tensor into one's favorite initial-value formulation
for the Einstein equations, as reviewed above, and implement the resulting
time evolution in a numerical scheme.

It is perhaps worth reflecting briefly on why we became interested
in revisiting this problem that has remained open for over forty years.
In earlier work \citep{Lowe:2022cne}, we studied how a semiclassical
limit emerges from a holographic description of quantum gravity such
as AdS/CFT. It quickly became apparent to us that while we might formulate
such a state in a well-defined way in conformal field theory variables,
it was not known quantitatively how a semiclassical state evolves
with respect to the gravity variables in the most interesting case
of four-dimensional spacetime. The purpose of the present work is
thus to develop a self-consistent set of dynamical equations for semiclassical
gravity coupled to matter in four dimensions. We have chosen a particularly
simple form of matter, and further simplification arises from taking
a large $N$ limit, but we expect our model to exhibit generic features
of semiclassical time evolution.

It is also worth commenting on what implications the present work
has for the information problem. There are many reasons to be cautious
in drawing conclusions based on previous results on semiclassical
gravity in four dimensions, which typically derive from computing
$\left\langle T_{\mu\nu}\right\rangle $ for matter is various quantum
states in a classical background geometry, which is either static
or patched together from different static solutions across shocks.
For instance, in the absence of semiclassical back-reaction, the combined
ADM mass of a black hole and its Hawking radiation diverges. So, there
is much to be gained from actually starting from a finite ADM mass
state and simply evolving it forward in time. Nevertheless, because
the evolution of at least the gravity variables is local, the resulting
solutions will exhibit smooth apparent horizons, allowing quantum
information to pass into the black hole interior. Therefore, at the
level of the semiclassical approximation, where gravity is essentially
treated as a unique classical field associated with some quantum matter
state, information will be lost. To go beyond this approximation requires
a fully quantum theory. If such a quantum theory shares the key features
of a holographic description such as AdS/CFT, then unitarity will
be preserved as described in \citep{Lowe:2022cne}.
\begin{acknowledgments}
Work supported in part by DOE grant de-sc0010010 Task A, Icelandic
Research Fund grant 228952-051, and a grant from the University of
Iceland Research Fund. D.L thanks M. Ronning, L. Brewin and A. Ori
for helpful discussions.
\end{acknowledgments}

\appendix

\section*{Appendix: Expansions of Subtraction Terms}

In this appendix, we provide short-distance expansions for various
quantities that enter in the point-splitting renormalization of the
scalar field stress-energy tensor. We begin by presenting the expansions
for the $U$ and $V$ terms, that appear in the subtraction term in
\eqref{eq:subtraction}, to the appropriate order in the coincident
limit. Since these functions are uniquely determined by the metric,
the expansions have been worked out long ago. Reference \citep{Decanini:2005eg}
gives\begin{dgroup*}
\[
U(x,x')=1+\frac{1}{12}R_{\mu\nu}\sigma^{;\mu}\sigma^{;\nu}-\frac{1}{24}R_{(\mu\nu;\lambda)}\sigma^{;\mu}\sigma^{;\nu}\sigma^{;\lambda}+\left(\frac{1}{80}R_{(\mu\nu;\lambda\rho)}+\frac{1}{288}R_{(\mu\nu}R_{\lambda\rho)}+\frac{1}{360}g_{\gamma\delta}R_{\alpha(\mu\,\,\nu}^{\,\,\,\,\gamma}R_{\,\,\lambda\,\,\rho)}^{\alpha\,\,\delta}\right)\sigma^{;\mu}\sigma^{;\nu}\sigma^{;\lambda}\sigma^{;\rho}+\mathcal{O}(\sigma^{5/2})
\]
\[
V(x,x')=\frac{1}{2}m^{2}+\frac{1}{2}\left(\xi-\frac{1}{6}\right)R-\frac{1}{4}\left(\xi-\frac{1}{6}\right)R_{;\mu}\sigma^{;\mu}+\left(\frac{1}{24}m^{2}R_{\mu\nu}+\frac{1}{12}\left(\xi-\frac{3}{20}\right)R_{;\mu\nu}-\frac{1}{240}\square R_{\mu\nu}+\frac{1}{24}\left(\xi-\frac{1}{6}\right)RR_{\mu\nu}+\frac{1}{180}R_{\,\mu}^{\alpha}R_{\alpha\nu}-\frac{1}{360}R^{\alpha\beta}R_{\alpha\mu\beta\nu}-\frac{1}{360}R_{\,\,\,\,\mu}^{\alpha\beta\gamma}R_{\alpha\beta\gamma\nu}\right)\sigma^{;\mu}\sigma^{;\nu}+\sigma\left(\frac{1}{8}m^{4}+\frac{1}{4}\left(\xi-\frac{1}{6}\right)m^{2}R-\frac{1}{24}\left(\xi-\frac{1}{5}\right)\square R+\frac{1}{8}\left(\xi-\frac{1}{6}\right)^{2}R^{2}-\frac{1}{720}R_{\mu\nu}R^{\mu\nu}+\frac{1}{720}R_{\mu\nu\lambda\rho}R^{\mu\nu\lambda\rho}\right)+\mathcal{O}(\sigma^{3/2})
\]
\end{dgroup*}The final element we need to compute $\left\langle T_{\mu\nu}^{bilocal}\right\rangle $
is an expression for $\sigma$ in the coincident limit. Numerical
methods are typically developed in a general coordinate system, rather
than a normal coordinate system. Hence our next task is to write expressions
for $\sigma$ and $\sigma^{;\mu}$ in general coordinates. This expansion
involves Christoffel symbol rather than curvature tensors. The Christoffel
symbols may be expressed in terms of the initial-value variables (whether
one chooses ADM or BSSN variables) and their spatial derivatives. 

The formula for $\sigma$ may be extracted from the expansions in
\citep{Brewin:2009se}. We will follow a similar approach and develop
the formulas using the Cadabra software package \citep{Peeters2018}.
In particular, if we define Riemann normal coordinates $y^{\mu}$
centered at the point $x$ then we have
\begin{equation}
\sigma=\frac{1}{2}L_{x,x'}^{2}=\frac{1}{2}g_{\mu\nu}(x)y_{x'}^{\mu}y_{x'}^{\nu},\label{eq:sigmaeqn}
\end{equation}
where $y_{x'}^{\mu}$ is the position of $x'$ in the patch of Riemann
normal coordinates near $x$. Then we need an expansion for $y_{x'}^{\mu}$
in terms of some general set of coordinates as $x'\to x,$ to fifth
order in the coordinate differences $\delta x^{\mu}=\left.x'\right.^{\mu}-x^{\mu}$,
\begin{equation}
y_{x'}^{\mu}=y^{(0)\mu}+y^{(1)\mu}+y^{(2)\mu}+y^{(3)\mu}+y^{(4)\mu},
\end{equation}
with\begin{dgroup*}
\[
y^{(0)\mu}=\delta x^{\mu}
\]
\[
y^{(1)\mu}=\frac{1}{2}\delta x^{\nu}\delta x^{\lambda}\Gamma_{\nu\lambda}^{\mu}
\]
\[
y^{(2)\mu}=\frac{1}{6}\delta x^{\nu}\delta x^{\lambda}\delta x^{\rho}\left(\Gamma_{\nu\tau}^{\mu}\Gamma_{\lambda\rho}^{\tau}+\partial_{\nu}\Gamma_{\lambda\rho}^{\mu}\right)
\]
\[
y^{(3)\mu}=\frac{1}{24}\delta x^{\nu}\delta x^{\lambda}\delta x^{\rho}\delta x^{\tau}\left(2\Gamma_{\nu\kappa}^{\mu}\partial_{\lambda}\Gamma_{\rho\tau}^{\kappa}+\Gamma_{\kappa\eta}^{\mu}\Gamma_{\nu\lambda}^{\kappa}\Gamma_{\rho\tau}^{\eta}+\Gamma_{\nu\lambda}^{\kappa}\partial_{\kappa}\Gamma_{\rho\tau}^{\mu}+\partial_{\nu}\partial_{\lambda}\Gamma_{\rho\tau}^{\mu}\right)
\]
\[
y^{(4)\mu}=\frac{1}{360}\delta x^{\nu}\delta x^{\lambda}\delta x^{\rho}\delta x^{\tau}\delta x^{\kappa}\left(-4\Gamma_{\nu\eta}^{\mu}\Gamma_{\lambda\zeta}^{\eta}\Gamma_{\rho\epsilon}^{\zeta}\Gamma_{\tau\kappa}^{\epsilon}+2\Gamma_{\nu\eta}^{\mu}\Gamma_{\lambda\zeta}^{\eta}\partial_{\rho}\Gamma_{\tau\kappa}^{\zeta}+3\Gamma_{\nu\eta}^{\mu}\Gamma_{\zeta\epsilon}^{\eta}\Gamma_{\lambda\rho}^{\zeta}\Gamma_{\tau\kappa}^{\epsilon}-6\Gamma_{\nu\eta}^{\mu}\Gamma_{\lambda\rho}^{\zeta}\partial_{\tau}\Gamma_{\kappa\zeta}^{\eta}+6\Gamma_{\nu\eta}^{\mu}\Gamma_{\lambda\rho}^{\zeta}\partial_{\zeta}\Gamma_{\tau\kappa}^{\eta}+9\Gamma_{\nu\eta}^{\mu}\partial_{\lambda\rho}\Gamma_{\tau\kappa}^{\eta}+4\Gamma_{\eta\zeta}^{\mu}\Gamma_{\nu\lambda}^{\eta}\Gamma_{\rho\epsilon}^{\zeta}\Gamma_{\tau\kappa}^{\epsilon}+13\Gamma_{\eta\zeta}^{\mu}\Gamma_{\nu\lambda}^{\eta}\partial_{\rho}\Gamma_{\tau\kappa}^{\zeta}-4\Gamma_{\nu\lambda}^{\eta}\Gamma_{\rho\eta}^{\zeta}\partial_{\tau}\Gamma_{\kappa\zeta}^{\mu}+\Gamma_{\nu\lambda}^{\eta}\Gamma_{\rho\eta}^{\zeta}\partial_{\zeta}\Gamma_{\tau\kappa}^{\mu}+2\left(\partial_{\nu}\Gamma_{\lambda\eta}^{\mu}\right)\left(\partial_{\rho}\Gamma_{\tau\kappa}^{\eta}\right)++7\left(\partial_{\eta}\Gamma_{\nu\lambda}^{\mu}\right)\left(\partial_{\rho}\Gamma_{\tau\kappa}^{\eta}\right)+3\Gamma_{\nu\lambda}^{\eta}\Gamma_{\rho\tau}^{\zeta}\partial_{\kappa}\Gamma_{\eta\zeta}^{\mu}+3\Gamma_{\nu\lambda}^{\eta}\Gamma_{\rho\tau}^{\zeta}\partial_{\eta}\Gamma_{\kappa\zeta}^{\mu}-3\Gamma_{\nu\lambda}^{\eta}\partial_{\rho}\partial_{\tau}\Gamma_{\kappa\eta}^{\mu}+6\Gamma_{\nu\lambda}^{\eta}\partial_{\rho}\partial_{\eta}\Gamma_{\tau\kappa}^{\mu}+3\partial_{\nu}\partial_{\lambda}\partial_{\rho}\Gamma_{\tau\kappa}^{\mu}\right)
\]
\end{dgroup*}which allows us to construct \eqref{eq:sigmaeqn} to
sixth order in the $\delta x^{\mu}$ as required
\begin{equation}
\sigma=\sigma^{(2)}+\sigma^{(3)}+\sigma^{(4)}+\sigma^{(5)}+\sigma^{(6)}+\mathcal{O}(\sigma^{7/2})
\end{equation}
\begin{dgroup*}
\[
\sigma^{(2)}=\frac{1}{2}\delta x^{\mu}\delta x^{\nu}g_{\mu\nu}
\]
\[
\sigma^{(3)}=\frac{1}{2}\delta x^{\mu}\delta x^{\nu}\delta x^{\alpha}g_{\mu\lambda}\Gamma^{\lambda}\,_{\nu\alpha}
\]
\[
\sigma^{(4)}=\frac{1}{24}\delta x^{\mu}\delta x^{\nu}\delta x^{\alpha}\delta x^{\lambda}\left(4g_{\mu\rho}\Gamma^{\rho}\,_{\nu\tau}\Gamma^{\tau}\,_{\alpha\lambda}+4g_{\mu\rho}\partial_{\nu}{\Gamma^{\rho}\,_{\alpha\lambda}}+3g_{\rho\tau}\Gamma^{\rho}\,_{\mu\nu}\Gamma^{\tau}\,_{\alpha\lambda}\right)
\]
\[
\sigma^{(5)}=\frac{1}{24}\delta x^{\mu}\delta x^{\nu}\delta x^{\alpha}\delta x^{\lambda}\delta x^{\rho}\left(2g_{\mu\tau}\Gamma^{\tau}\,_{\nu\kappa}\partial_{\alpha}{\Gamma^{\kappa}\,_{\lambda\rho}}+g_{\mu\tau}\Gamma^{\tau}\,_{\kappa\epsilon}\Gamma^{\kappa}\,_{\nu\alpha}\Gamma^{\epsilon}\,_{\lambda\rho}+g_{\mu\tau}\Gamma^{\kappa}\,_{\nu\alpha}\partial_{\kappa}{\Gamma^{\tau}\,_{\lambda\rho}}+g_{\mu\tau}\partial_{\nu}\partial_{\alpha}{\Gamma^{\tau}\,_{\lambda\rho}}+2g_{\tau\kappa}\Gamma^{\tau}\,_{\mu\nu}\Gamma^{\kappa}\,_{\alpha\epsilon}\Gamma^{\epsilon}\,_{\lambda\rho}+2g_{\tau\kappa}\Gamma^{\tau}\,_{\mu\nu}\partial_{\alpha}{\Gamma^{\kappa}\,_{\lambda\rho}}\right)
\]
\[
\sigma^{(6)}=\frac{1}{720}\delta x^{\mu}\delta x^{\nu}\delta x^{\alpha}\delta x^{\lambda}\delta x^{\rho}\delta x^{\tau}\left(-8g_{\mu\kappa}\Gamma^{\kappa}\,_{\nu\epsilon}\Gamma^{\epsilon}\,_{\alpha\eta}\Gamma^{\eta}\,_{\lambda\xi}\Gamma^{\xi}\,_{\rho\tau}+4g_{\mu\kappa}\Gamma^{\kappa}\,_{\nu\epsilon}\Gamma^{\epsilon}\,_{\alpha\eta}\partial_{\lambda}{\Gamma^{\eta}\,_{\rho\tau}}+6g_{\mu\kappa}\Gamma^{\kappa}\,_{\nu\epsilon}\Gamma^{\epsilon}\,_{\eta\xi}\Gamma^{\eta}\,_{\alpha\lambda}\Gamma^{\xi}\,_{\rho\tau}-12g_{\mu\kappa}\Gamma^{\kappa}\,_{\nu\epsilon}\Gamma^{\eta}\,_{\alpha\lambda}\partial_{\rho}{\Gamma^{\epsilon}\,_{\tau\eta}}+12g_{\mu\kappa}\Gamma^{\kappa}\,_{\nu\epsilon}\Gamma^{\eta}\,_{\alpha\lambda}\partial_{\eta}{\Gamma^{\epsilon}\,_{\rho\tau}}+18g_{\mu\kappa}\Gamma^{\kappa}\,_{\nu\epsilon}\partial_{\alpha}\partial_{\lambda}{\Gamma^{\epsilon}\,_{\rho\tau}}+8g_{\mu\kappa}\Gamma^{\kappa}\,_{\epsilon\eta}\Gamma^{\epsilon}\,_{\nu\alpha}\Gamma^{\eta}\,_{\lambda\xi}\Gamma^{\xi}\,_{\rho\tau}+26g_{\mu\kappa}\Gamma^{\kappa}\,_{\epsilon\eta}\Gamma^{\epsilon}\,_{\nu\alpha}\partial_{\lambda}{\Gamma^{\eta}\,_{\rho\tau}}-8g_{\mu\kappa}\Gamma^{\epsilon}\,_{\nu\alpha}\Gamma^{\eta}\,_{\lambda\epsilon}\partial_{\rho}{\Gamma^{\kappa}\,_{\tau\eta}}+2g_{\mu\kappa}\Gamma^{\epsilon}\,_{\nu\alpha}\Gamma^{\eta}\,_{\lambda\epsilon}\partial_{\eta}{\Gamma^{\kappa}\,_{\rho\tau}}+4g_{\mu\kappa}\partial_{\nu}{\Gamma^{\kappa}\,_{\alpha\epsilon}}\partial_{\lambda}{\Gamma^{\epsilon}\,_{\rho\tau}}+14g_{\mu\kappa}\partial_{\nu}{\Gamma^{\epsilon}\,_{\alpha\lambda}}\partial_{\epsilon}{\Gamma^{\kappa}\,_{\rho\tau}}+6g_{\mu\kappa}\Gamma^{\epsilon}\,_{\nu\alpha}\Gamma^{\eta}\,_{\lambda\rho}\partial_{\tau}{\Gamma^{\kappa}\,_{\epsilon\eta}}+6g_{\mu\kappa}\Gamma^{\epsilon}\,_{\nu\alpha}\Gamma^{\eta}\,_{\lambda\rho}\partial_{\epsilon}{\Gamma^{\kappa}\,_{\tau\eta}}-6g_{\mu\kappa}\Gamma^{\epsilon}\,_{\nu\alpha}\partial_{\lambda}\partial_{\rho}{\Gamma^{\kappa}\,_{\tau\epsilon}}+12g_{\mu\kappa}\Gamma^{\epsilon}\,_{\nu\alpha}\partial_{\lambda}\partial_{\epsilon}{\Gamma^{\kappa}\,_{\rho\tau}}+6g_{\mu\kappa}\partial_{\nu}\partial_{\alpha}\partial_{\lambda}{\Gamma^{\kappa}\,_{\rho\tau}}+30g_{\kappa\epsilon}\Gamma^{\kappa}\,_{\mu\nu}\Gamma^{\epsilon}\,_{\alpha\eta}\partial_{\lambda}{\Gamma^{\eta}\,_{\rho\tau}}+15g_{\kappa\epsilon}\Gamma^{\kappa}\,_{\mu\nu}\Gamma^{\epsilon}\,_{\eta\xi}\Gamma^{\eta}\,_{\alpha\lambda}\Gamma^{\xi}\,_{\rho\tau}%
+15g_{\kappa\epsilon}\Gamma^{\kappa}\,_{\mu\nu}\Gamma^{\eta}\,_{\alpha\lambda}\partial_{\eta}{\Gamma^{\epsilon}\,_{\rho\tau}}+15g_{\kappa\epsilon}\Gamma^{\kappa}\,_{\mu\nu}\partial_{\alpha}\partial_{\lambda}{\Gamma^{\epsilon}\,_{\rho\tau}}+10g_{\kappa\epsilon}\Gamma^{\kappa}\,_{\mu\eta}\Gamma^{\epsilon}\,_{\nu\xi}\Gamma^{\eta}\,_{\alpha\lambda}\Gamma^{\xi}\,_{\rho\tau}+20g_{\kappa\epsilon}\Gamma^{\kappa}\,_{\mu\eta}\Gamma^{\eta}\,_{\nu\alpha}\partial_{\lambda}{\Gamma^{\epsilon}\,_{\rho\tau}}+10g_{\kappa\epsilon}\partial_{\mu}{\Gamma^{\kappa}\,_{\nu\alpha}}\partial_{\lambda}{\Gamma^{\epsilon}\,_{\rho\tau}}\right)
\]
\end{dgroup*} Note that all the connection terms are evaluated at
the point $x$. We also need
\begin{equation}
\sigma^{;\mu}=g^{\mu\nu}\frac{\partial\sigma}{\partial x^{\nu}}
\end{equation}
out to fourth order in $\delta x^{\mu}$ which may be computed by
direct differentiation of \eqref{eq:sigmaeqn}. The result is
\begin{equation}
\sigma^{;\mu}=\sigma^{;\mu(1)}+\sigma^{;\mu(2)}+\sigma^{;\mu(3)}+\sigma^{;\mu(4)}+\mathcal{O}(\sigma^{5/2})
\end{equation}
\begin{dgroup*}
\[
\sigma^{;\mu(1)}=-\delta x^{\mu}
\]
\[
\sigma^{;\mu(2)}=-\frac{1}{2}\delta x^{\nu}\delta x^{\alpha}\Gamma^{\mu}\,_{\nu\alpha}
\]
\[
\sigma^{;\mu(3)}=\frac{1}{6}\delta x^{\nu}\delta x^{\alpha}\delta x^{\lambda}\left(2g^{\mu\rho}g_{\nu\tau}\Gamma^{\tau}\,_{\rho\kappa}\Gamma^{\kappa}\,_{\alpha\lambda}+2g^{\mu\rho}g_{\nu\tau}\partial_{\rho}{\Gamma^{\tau}\,_{\alpha\lambda}}-\Gamma^{\mu}\,_{\nu\rho}\Gamma^{\rho}\,_{\alpha\lambda}-2g^{\mu\rho}g_{\nu\tau}\Gamma^{\tau}\,_{\alpha\kappa}\Gamma^{\kappa}\,_{\lambda\rho}-\partial_{\nu}{\Gamma^{\mu}\,_{\alpha\lambda}}-2g^{\mu\rho}g_{\nu\tau}\partial_{\alpha}{\Gamma^{\tau}\,_{\lambda\rho}}\right)
\]
\[
\sigma^{;\mu(4)}=\frac{1}{24}\delta x^{\nu}\delta x^{\alpha}\delta x^{\lambda}\delta x^{\rho}\left(4g^{\mu\tau}g_{\nu\kappa}\Gamma^{\kappa}\,_{\tau\epsilon}\Gamma^{\epsilon}\,_{\alpha\eta}\Gamma^{\eta}\,_{\lambda\rho}+4g^{\mu\tau}g_{\nu\kappa}\Gamma^{\epsilon}\,_{\alpha\lambda}\partial_{\tau}{\Gamma^{\kappa}\,_{\rho\epsilon}}+2g^{\mu\tau}g_{\nu\kappa}\Gamma^{\kappa}\,_{\alpha\epsilon}\partial_{\tau}{\Gamma^{\epsilon}\,_{\lambda\rho}}+2g^{\mu\tau}g_{\nu\kappa}\Gamma^{\kappa}\,_{\tau\epsilon}\partial_{\alpha}{\Gamma^{\epsilon}\,_{\lambda\rho}}+2g^{\mu\tau}g_{\nu\kappa}\partial_{\alpha}\partial_{\tau}{\Gamma^{\kappa}\,_{\lambda\rho}}+4g^{\mu\tau}g_{\kappa\epsilon}\Gamma^{\kappa}\,_{\nu\alpha}\Gamma^{\epsilon}\,_{\tau\eta}\Gamma^{\eta}\,_{\lambda\rho}+4g^{\mu\tau}g_{\kappa\epsilon}\Gamma^{\kappa}\,_{\nu\alpha}\partial_{\tau}{\Gamma^{\epsilon}\,_{\lambda\rho}}-2\Gamma^{\mu}\,_{\nu\tau}\partial_{\alpha}{\Gamma^{\tau}\,_{\lambda\rho}}-4g^{\mu\tau}g_{\nu\kappa}\Gamma^{\kappa}\,_{\alpha\epsilon}\partial_{\lambda}{\Gamma^{\epsilon}\,_{\rho\tau}}-\Gamma^{\mu}\,_{\tau\kappa}\Gamma^{\tau}\,_{\nu\alpha}\Gamma^{\kappa}\,_{\lambda\rho}-4g^{\mu\tau}g_{\nu\kappa}\Gamma^{\kappa}\,_{\epsilon\eta}\Gamma^{\epsilon}\,_{\alpha\lambda}\Gamma^{\eta}\,_{\rho\tau}-\Gamma^{\tau}\,_{\nu\alpha}\partial_{\tau}{\Gamma^{\mu}\,_{\lambda\rho}}-2g^{\mu\tau}g_{\nu\kappa}\Gamma^{\epsilon}\,_{\alpha\tau}\partial_{\epsilon}{\Gamma^{\kappa}\,_{\lambda\rho}}-2g^{\mu\tau}g_{\nu\kappa}\Gamma^{\epsilon}\,_{\alpha\lambda}\partial_{\epsilon}{\Gamma^{\kappa}\,_{\rho\tau}}-\partial_{\nu}\partial_{\alpha}{\Gamma^{\mu}\,_{\lambda\rho}}-2g^{\mu\tau}g_{\nu\kappa}\partial_{\alpha}\partial_{\lambda}{\Gamma^{\kappa}\,_{\rho\tau}}-4g^{\mu\tau}g_{\kappa\epsilon}\Gamma^{\kappa}\,_{\nu\alpha}\Gamma^{\epsilon}\,_{\lambda\eta}\Gamma^{\eta}\,_{\rho\tau}-4g^{\mu\tau}g_{\kappa\epsilon}\Gamma^{\kappa}\,_{\nu\alpha}\partial_{\lambda}{\Gamma^{\epsilon}\,_{\rho\tau}}\right)
\]
\end{dgroup*} Finally we need the bivector of parallel transport,
also expanded out to fourth order in $\delta x^{\mu}$. We obtain
this by solving the equation $\sigma^{;\mu}\nabla_{\mu}\bar{g}^{\alpha\beta'}=0$.
\[
\bar{g}^{\alpha\beta'}=g^{\alpha\beta}+\bar{g}^{\alpha\beta(1)}+\bar{g}^{\alpha\beta(2)}+\bar{g}^{\alpha\beta(3)}+\bar{g}^{\alpha\beta(4)}+\cdots
\]
 \begin{dgroup*}
\[
\bar{g}^{\alpha\beta(1)}=-\delta x^{\mu}g^{\alpha\nu}\Gamma^{\beta}\,_{\mu\nu}
\]
\[
\bar{g}^{\alpha\beta(2)}=\frac{1}{2}\delta x^{\mu}\delta x^{\nu}\left(g^{\alpha\lambda}\Gamma^{\beta}\,_{\mu\rho}\Gamma^{\rho}\,_{\nu\lambda}-g^{\alpha\lambda}\partial_{\mu}{\Gamma^{\beta}\,_{\nu\lambda}}\right)
\]
\[
\bar{g}^{\alpha\beta(3)}=\frac{1}{18}\delta x^{\mu}\delta x^{\nu}\delta x^{\lambda}\left(-3g^{\alpha\rho}\Gamma^{\beta}\,_{\mu\tau}\Gamma^{\tau}\,_{\nu\kappa}\Gamma^{\kappa}\,_{\lambda\rho}+6g^{\alpha\rho}\Gamma^{\tau}\,_{\mu\rho}\partial_{\nu}{\Gamma^{\beta}\,_{\lambda\tau}}+3g^{\alpha\rho}\Gamma^{\beta}\,_{\mu\tau}\partial_{\nu}{\Gamma^{\tau}\,_{\lambda\rho}}-3g^{\alpha\rho}\partial_{\mu}\partial_{\nu}{\Gamma^{\beta}\,_{\lambda\rho}}-3g^{\alpha\rho}\Gamma^{\tau}\,_{\mu\nu}\partial_{\tau}{\Gamma^{\beta}\,_{\lambda\rho}}+3g^{\alpha\rho}\Gamma^{\tau}\,_{\mu\nu}\partial_{\lambda}{\Gamma^{\beta}\,_{\rho\tau}}\right)
\]
\[
\bar{g}^{\alpha\beta(4)}=\frac{1}{144}\delta x^{\mu}\delta x^{\nu}\delta x^{\lambda}\delta x^{\rho}\left(6g^{\alpha\tau}\Gamma^{\beta}\,_{\mu\kappa}\Gamma^{\kappa}\,_{\nu\epsilon}\Gamma^{\epsilon}\,_{\lambda\eta}\Gamma^{\eta}\,_{\rho\tau}-18g^{\alpha\tau}\Gamma^{\kappa}\,_{\mu\tau}\Gamma^{\epsilon}\,_{\nu\kappa}\partial_{\lambda}{\Gamma^{\beta}\,_{\rho\epsilon}}-12g^{\alpha\tau}\Gamma^{\beta}\,_{\mu\kappa}\Gamma^{\epsilon}\,_{\nu\tau}\partial_{\lambda}{\Gamma^{\kappa}\,_{\rho\epsilon}}-6g^{\alpha\tau}\Gamma^{\beta}\,_{\mu\kappa}\Gamma^{\kappa}\,_{\nu\epsilon}\partial_{\lambda}{\Gamma^{\epsilon}\,_{\rho\tau}}+18g^{\alpha\tau}\partial_{\mu}{\Gamma^{\beta}\,_{\nu\kappa}}\partial_{\lambda}{\Gamma^{\kappa}\,_{\rho\tau}}+18g^{\alpha\tau}\Gamma^{\kappa}\,_{\mu\tau}\partial_{\nu}\partial_{\lambda}{\Gamma^{\beta}\,_{\rho\kappa}}+6g^{\alpha\tau}\Gamma^{\beta}\,_{\mu\kappa}\partial_{\nu}\partial_{\lambda}{\Gamma^{\kappa}\,_{\rho\tau}}-6g^{\alpha\tau}\partial_{\mu}\partial_{\nu}\partial_{\lambda}{\Gamma^{\beta}\,_{\rho\tau}}+15g^{\alpha\tau}\Gamma^{\kappa}\,_{\mu\nu}\Gamma^{\epsilon}\,_{\lambda\tau}\partial_{\kappa}{\Gamma^{\beta}\,_{\rho\epsilon}}-12g^{\alpha\tau}\partial_{\kappa}{\Gamma^{\beta}\,_{\mu\tau}}\partial_{\nu}{\Gamma^{\kappa}\,_{\lambda\rho}}-15g^{\alpha\tau}\Gamma^{\kappa}\,_{\mu\nu}\partial_{\lambda}\partial_{\kappa}{\Gamma^{\beta}\,_{\rho\tau}}-24g^{\alpha\tau}\Gamma^{\kappa}\,_{\mu\nu}\Gamma^{\epsilon}\,_{\lambda\tau}\partial_{\rho}{\Gamma^{\beta}\,_{\kappa\epsilon}}+12g^{\alpha\tau}\partial_{\mu}{\Gamma^{\beta}\,_{\tau\kappa}}\partial_{\nu}{\Gamma^{\kappa}\,_{\lambda\rho}}+15g^{\alpha\tau}\Gamma^{\kappa}\,_{\mu\nu}\partial_{\lambda}\partial_{\rho}{\Gamma^{\beta}\,_{\tau\kappa}}+9g^{\alpha\tau}\Gamma^{\beta}\,_{\mu\kappa}\Gamma^{\epsilon}\,_{\nu\lambda}\partial_{\epsilon}{\Gamma^{\kappa}\,_{\rho\tau}}+9g^{\alpha\tau}\Gamma^{\kappa}\,_{\mu\nu}\Gamma^{\epsilon}\,_{\tau\kappa}\partial_{\lambda}{\Gamma^{\beta}\,_{\rho\epsilon}}-9g^{\alpha\tau}\Gamma^{\beta}\,_{\mu\kappa}\Gamma^{\epsilon}\,_{\nu\lambda}\partial_{\rho}{\Gamma^{\kappa}\,_{\tau\epsilon}}+12g^{\alpha\tau}g^{\kappa\epsilon}g_{\mu\eta}\Gamma^{\eta}\,_{\kappa\xi}\Gamma^{\xi}\,_{\nu\lambda}\partial_{\epsilon}{\Gamma^{\beta}\,_{\rho\tau}}-12g^{\alpha\tau}g^{\kappa\epsilon}g_{\mu\eta}\Gamma^{\eta}\,_{\kappa\xi}\Gamma^{\xi}\,_{\nu\lambda}\partial_{\rho}{\Gamma^{\beta}\,_{\tau\epsilon}}%
+12g^{\alpha\tau}g^{\kappa\epsilon}g_{\mu\eta}\partial_{\kappa}{\Gamma^{\beta}\,_{\nu\tau}}\partial_{\epsilon}{\Gamma^{\eta}\,_{\lambda\rho}}-12g^{\alpha\tau}g^{\kappa\epsilon}g_{\mu\eta}\partial_{\nu}{\Gamma^{\beta}\,_{\tau\kappa}}\partial_{\epsilon}{\Gamma^{\eta}\,_{\lambda\rho}}-6g^{\alpha\tau}\Gamma^{\kappa}\,_{\mu\nu}\Gamma^{\epsilon}\,_{\lambda\kappa}\partial_{\epsilon}{\Gamma^{\beta}\,_{\rho\tau}}+6g^{\alpha\tau}\Gamma^{\kappa}\,_{\mu\nu}\Gamma^{\epsilon}\,_{\lambda\kappa}\partial_{\rho}{\Gamma^{\beta}\,_{\tau\epsilon}}-12g^{\alpha\tau}g^{\kappa\epsilon}g_{\mu\eta}\Gamma^{\eta}\,_{\nu\xi}\Gamma^{\xi}\,_{\lambda\kappa}\partial_{\epsilon}{\Gamma^{\beta}\,_{\rho\tau}}+12g^{\alpha\tau}g^{\kappa\epsilon}g_{\mu\eta}\Gamma^{\eta}\,_{\nu\xi}\Gamma^{\xi}\,_{\lambda\kappa}\partial_{\rho}{\Gamma^{\beta}\,_{\tau\epsilon}}-12g^{\alpha\tau}g^{\kappa\epsilon}g_{\mu\eta}\partial_{\kappa}{\Gamma^{\beta}\,_{\nu\tau}}\partial_{\lambda}{\Gamma^{\eta}\,_{\rho\epsilon}}+12g^{\alpha\tau}g^{\kappa\epsilon}g_{\mu\eta}\partial_{\nu}{\Gamma^{\beta}\,_{\tau\kappa}}\partial_{\lambda}{\Gamma^{\eta}\,_{\rho\epsilon}}\right)
\]
\end{dgroup*}

\bibliographystyle{utcaps}
\bibliography{hawking}

\providecommand{\href}[2]{#2}\begingroup\raggedright\begin{thebibliography}{10}

\bibitem{Hawking:1975vcx}
S.~W. Hawking, ``{Particle Creation by Black Holes},''
  \href{http://dx.doi.org/10.1007/BF02345020}{{\em Commun. Math. Phys.}
  {\bfseries 43} (1975) 199--220}. [Erratum: Commun.Math.Phys. 46, 206 (1976)].

\bibitem{Birrell:1982ix}
N.~D. Birrell and P.~C.~W. Davies,
  \href{http://dx.doi.org/10.1017/CBO9780511622632}{{\em {Quantum Fields in
  Curved Space}}}.
\newblock Cambridge Monographs on Mathematical Physics. Cambridge Univ. Press,
  Cambridge, UK, 2, 1984.

\bibitem{Wald:1984rg}
R.~M. Wald,
  \href{http://dx.doi.org/10.7208/chicago/9780226870373.001.0001}{{\em {General
  Relativity}}}.
\newblock Chicago Univ. Pr., Chicago, USA, 1984.

\bibitem{Wald:1995yp}
R.~M. Wald, {\em {Quantum Field Theory in Curved Space-Time and Black Hole
  Thermodynamics}}.
\newblock Chicago Lectures in Physics. University of Chicago Press, Chicago,
  IL, 1995.

\bibitem{PhysRevD.14.870}
W.~G. Unruh, ``Notes on black-hole evaporation,''
  \href{http://dx.doi.org/10.1103/PhysRevD.14.870}{{\em Phys. Rev. D}
  {\bfseries 14} (Aug, 1976) 870--892}.
  \url{https://link.aps.org/doi/10.1103/PhysRevD.14.870}.

\bibitem{DeWitt:1960fc}
B.~S. DeWitt and R.~W. Brehme, ``{Radiation damping in a gravitational
  field},'' \href{http://dx.doi.org/10.1016/0003-4916(60)90030-0}{{\em Annals
  Phys.} {\bfseries 9} (1960) 220--259}.

\bibitem{DeWitt:1964mxt}
B.~S. DeWitt, ``{Dynamical theory of groups and fields},'' {\em Conf. Proc. C}
  {\bfseries 630701} (1964) 585--820.

\bibitem{Christensen:1976vb}
S.~M. Christensen, ``{Vacuum Expectation Value of the Stress Tensor in an
  Arbitrary Curved Background: The Covariant Point Separation Method},''
  \href{http://dx.doi.org/10.1103/PhysRevD.14.2490}{{\em Phys. Rev. D}
  {\bfseries 14} (1976) 2490--2501}.

\bibitem{Adler:1976jx}
S.~L. Adler, J.~Lieberman, and Y.~J. Ng, ``{Regularization of the Stress Energy
  Tensor for Vector and Scalar Particles Propagating in a General Background
  Metric},'' \href{http://dx.doi.org/10.1016/0003-4916(77)90313-X}{{\em Annals
  Phys.} {\bfseries 106} (1977) 279}.

\bibitem{Wald:1978pj}
R.~M. Wald, ``{Trace Anomaly of a Conformally Invariant Quantum Field in Curved
  Space-Time},'' \href{http://dx.doi.org/10.1103/PhysRevD.17.1477}{{\em Phys.
  Rev. D} {\bfseries 17} (1978) 1477--1484}.

\bibitem{Candelas:1980zt}
P.~Candelas, ``{Vacuum Polarization in Schwarzschild Space-Time},''
  \href{http://dx.doi.org/10.1103/PhysRevD.21.2185}{{\em Phys. Rev. D}
  {\bfseries 21} (1980) 2185--2202}.

\bibitem{Brown:1986tj}
M.~R. Brown and A.~C. Ottewill, ``{Photon Propagators and the Definition and
  Approximation of Renormalized Stress Tensors in Curved Space-time},''
  \href{http://dx.doi.org/10.1103/PhysRevD.34.1776}{{\em Phys. Rev. D}
  {\bfseries 34} (1986) 1776--1786}.

\bibitem{Decanini:2005eg}
Y.~Decanini and A.~Folacci, ``{Hadamard renormalization of the stress-energy
  tensor for a quantized scalar field in a general spacetime of arbitrary
  dimension},'' \href{http://dx.doi.org/10.1103/PhysRevD.78.044025}{{\em Phys.
  Rev. D} {\bfseries 78} (2008) 044025},
  \href{http://arxiv.org/abs/gr-qc/0512118}{{\ttfamily arXiv:gr-qc/0512118}}.

\bibitem{Barvinsky:1990up}
A.~O. Barvinsky and G.~A. Vilkovisky, ``{Covariant perturbation theory. 2:
  Second order in the curvature. General algorithms},''
  \href{http://dx.doi.org/10.1016/0550-3213(90)90047-H}{{\em Nucl. Phys. B}
  {\bfseries 333} (1990) 471--511}.

\bibitem{POLYAKOV1981207}
A.~Polyakov, ``Quantum geometry of bosonic strings,''
  \href{http://dx.doi.org/https://doi.org/10.1016/0370-2693(81)90743-7}{{\em
  Physics Letters B} {\bfseries 103} no.~3, (1981) 207--210}.
  \url{https://www.sciencedirect.com/science/article/pii/0370269381907437}.

\bibitem{Callan:1992rs}
C.~G. Callan, Jr., S.~B. Giddings, J.~A. Harvey, and A.~Strominger,
  ``{Evanescent black holes},''
  \href{http://dx.doi.org/10.1103/PhysRevD.45.R1005}{{\em Phys. Rev. D}
  {\bfseries 45} no.~4, (1992) R1005},
  \href{http://arxiv.org/abs/hep-th/9111056}{{\ttfamily arXiv:hep-th/9111056}}.

\bibitem{Calmet:2021cip}
X.~Calmet and S.~D.~H. Hsu, ``{Quantum hair and black hole information},''
  \href{http://dx.doi.org/10.1016/j.physletb.2022.136995}{{\em Phys. Lett. B}
  {\bfseries 827} (2022) 136995},
  \href{http://arxiv.org/abs/2112.05171}{{\ttfamily arXiv:2112.05171
  [hep-th]}}.

\bibitem{Levi:2015eea}
A.~Levi and A.~Ori, ``{Pragmatic mode-sum regularization method for
  semiclassical black-hole spacetimes},''
  \href{http://dx.doi.org/10.1103/PhysRevD.91.104028}{{\em Phys. Rev. D}
  {\bfseries 91} (2015) 104028},
  \href{http://arxiv.org/abs/1503.02810}{{\ttfamily arXiv:1503.02810 [gr-qc]}}.

\bibitem{Levi:2016quh}
A.~Levi and A.~Ori, ``{Versatile method for renormalized stress-energy
  computation in black-hole spacetimes},''
  \href{http://dx.doi.org/10.1103/PhysRevLett.117.231101}{{\em Phys. Rev.
  Lett.} {\bfseries 117} no.~23, (2016) 231101},
  \href{http://arxiv.org/abs/1608.03806}{{\ttfamily arXiv:1608.03806 [gr-qc]}}.

\bibitem{Zilberman:2019buh}
N.~Zilberman, A.~Levi, and A.~Ori, ``{Quantum fluxes at the inner horizon of a
  spherical charged black hole},''
  \href{http://dx.doi.org/10.1103/PhysRevLett.124.171302}{{\em Phys. Rev.
  Lett.} {\bfseries 124} no.~17, (2020) 171302},
  \href{http://arxiv.org/abs/1906.11303}{{\ttfamily arXiv:1906.11303 [gr-qc]}}.

\bibitem{Juarez-Aubry:2022qdp}
B.~A. Ju\'arez-Aubry, B.~S. Kay, T.~Miramontes, and D.~Sudarsky, ``{On the
  initial value problem for semiclassical gravity without and with quantum
  state collapses},''
  \href{http://dx.doi.org/10.1088/1475-7516/2023/01/040}{{\em JCAP} {\bfseries
  01} (2023) 040}, \href{http://arxiv.org/abs/2205.11671}{{\ttfamily
  arXiv:2205.11671 [gr-qc]}}.

\bibitem{Simon:1990jn}
J.~Z. Simon, ``{The Stability of flat space, semiclassical gravity, and higher
  derivatives},'' \href{http://dx.doi.org/10.1103/PhysRevD.43.3308}{{\em Phys.
  Rev. D} {\bfseries 43} (1991) 3308--3316}.

\bibitem{Parker:1993dk}
L.~Parker and J.~Z. Simon, ``{Einstein equation with quantum corrections
  reduced to second order},''
  \href{http://dx.doi.org/10.1103/PhysRevD.47.1339}{{\em Phys. Rev. D}
  {\bfseries 47} (1993) 1339--1355},
  \href{http://arxiv.org/abs/gr-qc/9211002}{{\ttfamily arXiv:gr-qc/9211002}}.

\bibitem{York:1978gql}
J.~W. York, Jr., ``{Kinematics and Dynamics of General Relativity},'' in {\em
  {Workshop on Sources of Gravitational Radiation}}, pp.~83--126.
\newblock 1978.

\bibitem{Gourgoulhon:2007ue}
E.~Gourgoulhon, ``{3+1 formalism and bases of numerical relativity},''
  \href{http://arxiv.org/abs/gr-qc/0703035}{{\ttfamily arXiv:gr-qc/0703035}}.

\bibitem{nakamura}
M.~Shibata and T.~Nakamura, ``Evolution of three-dimensional gravitational
  waves: Harmonic slicing case,''
  \href{http://dx.doi.org/10.1103/PhysRevD.52.5428}{{\em Phys. Rev. D}
  {\bfseries 52} (Nov, 1995) 5428--5444}.
  \url{https://link.aps.org/doi/10.1103/PhysRevD.52.5428}.

\bibitem{baumgarte}
T.~W. Baumgarte and S.~L. Shapiro, ``Numerical integration of Einstein's field
  equations,'' \href{http://dx.doi.org/10.1103/PhysRevD.59.024007}{{\em Phys.
  Rev. D} {\bfseries 59} (Dec, 1998) 024007}.
  \url{https://link.aps.org/doi/10.1103/PhysRevD.59.024007}.

\bibitem{Arnowitt:1962hi}
R.~L. Arnowitt, S.~Deser, and C.~W. Misner, ``{The Dynamics of General
  Relativity},'' \href{http://dx.doi.org/10.1007/s10714-008-0661-1}{{\em Gen.
  Rel. Grav.} {\bfseries 40} (2008) 1997--2027},
  \href{http://arxiv.org/abs/gr-qc/0405109}{{\ttfamily arXiv:gr-qc/0405109}}.

\bibitem{Russo:1992ax}
J.~G. Russo, L.~Susskind, and L.~Thorlacius, ``{The Endpoint of Hawking
  radiation},'' \href{http://dx.doi.org/10.1103/PhysRevD.46.3444}{{\em Phys.
  Rev. D} {\bfseries 46} (1992) 3444--3449},
  \href{http://arxiv.org/abs/hep-th/9206070}{{\ttfamily arXiv:hep-th/9206070}}.

\bibitem{Lowe:1992ed}
D.~A. Lowe, ``{Semiclassical approach to black hole evaporation},''
  \href{http://dx.doi.org/10.1103/PhysRevD.47.2446}{{\em Phys. Rev. D}
  {\bfseries 47} (1993) 2446--2453},
  \href{http://arxiv.org/abs/hep-th/9209008}{{\ttfamily arXiv:hep-th/9209008}}.

\bibitem{Lowe:1995ac}
D.~A. Lowe, J.~Polchinski, L.~Susskind, L.~Thorlacius, and J.~Uglum, ``{Black
  hole complementarity versus locality},''
  \href{http://dx.doi.org/10.1103/PhysRevD.52.6997}{{\em Phys. Rev. D}
  {\bfseries 52} (1995) 6997--7010},
  \href{http://arxiv.org/abs/hep-th/9506138}{{\ttfamily arXiv:hep-th/9506138}}.

\bibitem{Lowe:2022cne}
D.~A. Lowe and L.~Thorlacius, ``{Quantum chaos and unitary black hole
  evaporation},'' \href{http://dx.doi.org/10.1007/JHEP05(2022)165}{{\em JHEP}
  {\bfseries 05} (2022) 165}, \href{http://arxiv.org/abs/2203.06434}{{\ttfamily
  arXiv:2203.06434 [hep-th]}}.

\bibitem{candelas}
K.~W. Howard and P.~Candelas, ``Quantum Stress Tensor in Schwarzschild
  Space-Time,'' \href{http://dx.doi.org/10.1103/PhysRevLett.53.403}{{\em Phys.
  Rev. Lett.} {\bfseries 53} (Jul, 1984) 403--406}.
  \url{https://link.aps.org/doi/10.1103/PhysRevLett.53.403}.

\bibitem{Wald:1993nt}
R.~M. Wald, ``{Black hole entropy is the Noether charge},''
  \href{http://dx.doi.org/10.1103/PhysRevD.48.R3427}{{\em Phys. Rev. D}
  {\bfseries 48} no.~8, (1993) R3427--R3431},
  \href{http://arxiv.org/abs/gr-qc/9307038}{{\ttfamily arXiv:gr-qc/9307038}}.

\bibitem{Brewin:2009se}
L.~Brewin, ``{Riemann Normal Coordinate expansions using Cadabra},''
  \href{http://dx.doi.org/10.1088/0264-9381/26/17/175017}{{\em Class. Quant.
  Grav.} {\bfseries 26} (2009) 175017},
  \href{http://arxiv.org/abs/0903.2087}{{\ttfamily arXiv:0903.2087 [gr-qc]}}.

\bibitem{Peeters2018}
K.~Peeters, ``Cadabra2: computer algebra for field theory revisited,''
  \href{http://dx.doi.org/10.21105/joss.01118}{{\em Journal of Open Source
  Software} {\bfseries 3} no.~32, (2018) 1118}.
  \url{https://doi.org/10.21105/joss.01118}.

\end{thebibliography}\endgroup

\end{document}